\let\oldvec\vec
\documentclass[twocolumn,epjc3]{svjour3} 
\let\epjcvec\vec
\let\vec\oldvec

\RequirePackage[T1]{fontenc}
\usepackage[english]{babel}

\smartqed

\RequirePackage[pdftex]{graphicx}
\RequirePackage{mathptmx}
\RequirePackage{flushend}
\RequirePackage[numbers,sort&compress]{natbib}
\RequirePackage[colorlinks,citecolor=blue,urlcolor=blue,linkcolor=blue]{hyperref}
\usepackage{articlestyle}
\usepackage{DataMacros}
\let\vec\epjcvec

\journalname{Eur. Phys. J. C}

\begin{document}
\title{Heavy-Quarkonium Potential from the Lattice Gluon Propagator}

\author{Willian M. Serenone\thanksref{e1,addr1}
        \and
        Attilio Cucchieri\thanksref{e2,addr1}
        \and
        Tereza Mendes\thanksref{e3,addr1}
}

\thankstext{e1}{e-mail: willian.serenone@usp.br}
\thankstext{e2}{e-mail: attilio@ifsc.usp.br}
\thankstext{e3}{e-mail: mendes@ifsc.usp.br}

\institute{Instituto de F\'isica de S\~ao Carlos, Universidade de S\~ao Paulo \\
Caixa Postal 369, CEP 13560-970, S\~ao Carlos SP, Brazil\label{addr1}}

\date{Received: date / Accepted: date}

\maketitle

\begin{abstract}
We consider the potential-model approach for obtaining the spectrum of
charmonium and bottomonium, replacing the usual gluon propagator
by one obtained from lattice simulations. The resulting spectra are
compared to the corresponding ones in the Cornell-potential case. We
also estimate the interquark distance in both cases.
\end{abstract}

\section{Introduction}
\label{sec:Introduction}

A reliable description of heavy quarkonia states is of great interest for our understanding of 
nonperturbative aspects of QCD \cite{Brambilla:2014jmp} and is expected to be important 
in guiding the search for physics beyond the standard model \cite{Love:2008ys}.
A fortuitous advantage in the study of such states is that, due to the large mass of the 
heavy quarks, various approximations may be adopted. For example, an expansion in inverse 
powers of the heavy-quark mass $m$ is performed in potential nonrelativistic QCD (pNRQCD) 
\cite{Brambilla:2004jw}, and lattice simulations (especially for bottomonium systems) are 
applied to effective actions obtained by an expansion in powers of the heavy-quark velocity $v/c$.
Similarly, in the relativistic quark model with the quasipotential approach, radiative corrections may
be included and treated perturbatively in the case of heavy quarkonia \cite{Ebert:2011jc}.
This possibility of exploring different scales of the problem separately is also
helpful in methods more directly based on QCD, such as studies of Dyson-Schwinger and Bethe-Salpeter
equations \cite{Fischer:2014cfa}. 

An early but still successful approach to describe heavy quarkonia is given by nonrelativistic 
potential models, to which relativistic corrections may also be added \cite{Radford:2007vd}.
\footnote{Note that these corrections may be computed from lattice data for the Wilson loop \cite{Koma:2006si,Koma:2008zza}.}
The idea is to view confinement as an ``a priori'' property of QCD, modeling
the interquark potential to incorporate some known features of the interaction at both
ends of the energy scale.
The simplest such model, the Cornell --- or Coulomb-plus-linear --- potential
\cite{Eichten:1978tg,Eichten:1979ms,Eichten:2002qv}, is 
obtained by supplementing the high-energy (perturbative) part of the potential with an explicit 
confining term. The resulting expression is a sum of two terms: the first one comes from the 
quark-antiquark interaction in the one-gluon-exchange (OGE) approximation  
and the second one is a linearly rising potential.
We have
\begin{equation}
\label{Cornell}
V(r)\;=\; -\frac{4}{3}\frac{\alpha_s}{r} \,+\, \sigma\,r\,,
\end{equation}
where ${\alpha_s}$ is the strong coupling constant and $\sigma$ is the string tension.
The first term may be associated with scattering of the quark-antiquark pair inside the meson and
is analogous to the Coulomb potential in the QED case. The second term corresponds to linear 
confinement as observed from the strong-coupling expansion of the Wilson loop in lattice gauge theory
with static quarks.

In practice, the static interquark potential may be defined conveniently in terms of the
Wilson loop, or it may be obtained (perturbatively) by taking the nonrelativistic limit in the Bethe-Salpeter 
equation describing the bound state of two heavy fermions. This yields a Schr\"odinger equation, to which
a linear term is added a posteriori. 
The numerical procedure for obtaining the mass spectrum for the 
Cornell potential, as well as for other commonly used potentials is reviewed in 
detail in \cite{Lucha:1991vn}.

The Cornell potential provides a spin-independent description of
the interquark potential for heavy quarks, with parameters determined by fitting a few 
known states (see e.g.\ \cite{Chung:2008sm}) or by comparison with lattice simulations.
For a recent determination of these parameters, see Ref.\ \cite{Kawanai:2011xb}.
It would be interesting, nevertheless, to have a better insight about confinement as an emergent 
property of the interquark interaction induced by the gluon propagator, rather than as a 
built-in feature.

Of course, the gluon propagator in QCD is very different from the perturbative one
at the hadronic scale, and it should contain full information about confinement.
In order to use this nonperturbative information we propose to substitute the free gluon 
propagator in the OGE term of the potential, as described above, by a fully nonperturbative one, 
obtained from lattice simulations. We want to check if this replacement leads to an improved 
description of the spectra, possibly without the need to include the linearly rising term explicitly.
We use the data generated in studies of the SU(2) gluon
propagator in Landau gauge on very large lattices (up to $128^4$), reported in
\cite{Cucchieri:2007md,Cucchieri:2007rg}.
More precisely, we use directly the fit obtained in Ref.\ \cite{Cucchieri:2011ig}.
We note that our aim is to gain a qualitative understanding of the interplay between perturbative
and nonperturbative features of the interquark potential.
Our approach is similar in spirit to the one in Refs.\ \cite{Gonzalez:2011zc,Vento:2012wp},
but our conclusions are different.

Preliminary versions of our study have been presented in Refs.\ \cite{Serenone:2012yta} and \cite{Serenone:2014ota}.

We organize this paper in the following way. In Section \ref{sec:Pot_Model_Review} we review
the procedure for obtaining the Coulomb potential in QED as the nonrelativistic limit of $e^- e^+$ 
scattering and how this is adapted to heavy quarks. In particular, we introduce the lattice 
propagator in the OGE term. 
In Section \ref{sec:Methods} we describe our method for obtaining the mass spectra with the
desired potential. Our results are presented in Section \ref{sec:Results} and our conclusions
in Section \ref{sec:Conclusions}.

\section{Potential from Lattice Propagator}
\label{sec:Pot_Model_Review}

\begin{figure}%
\includegraphics[width=0.6\columnwidth]{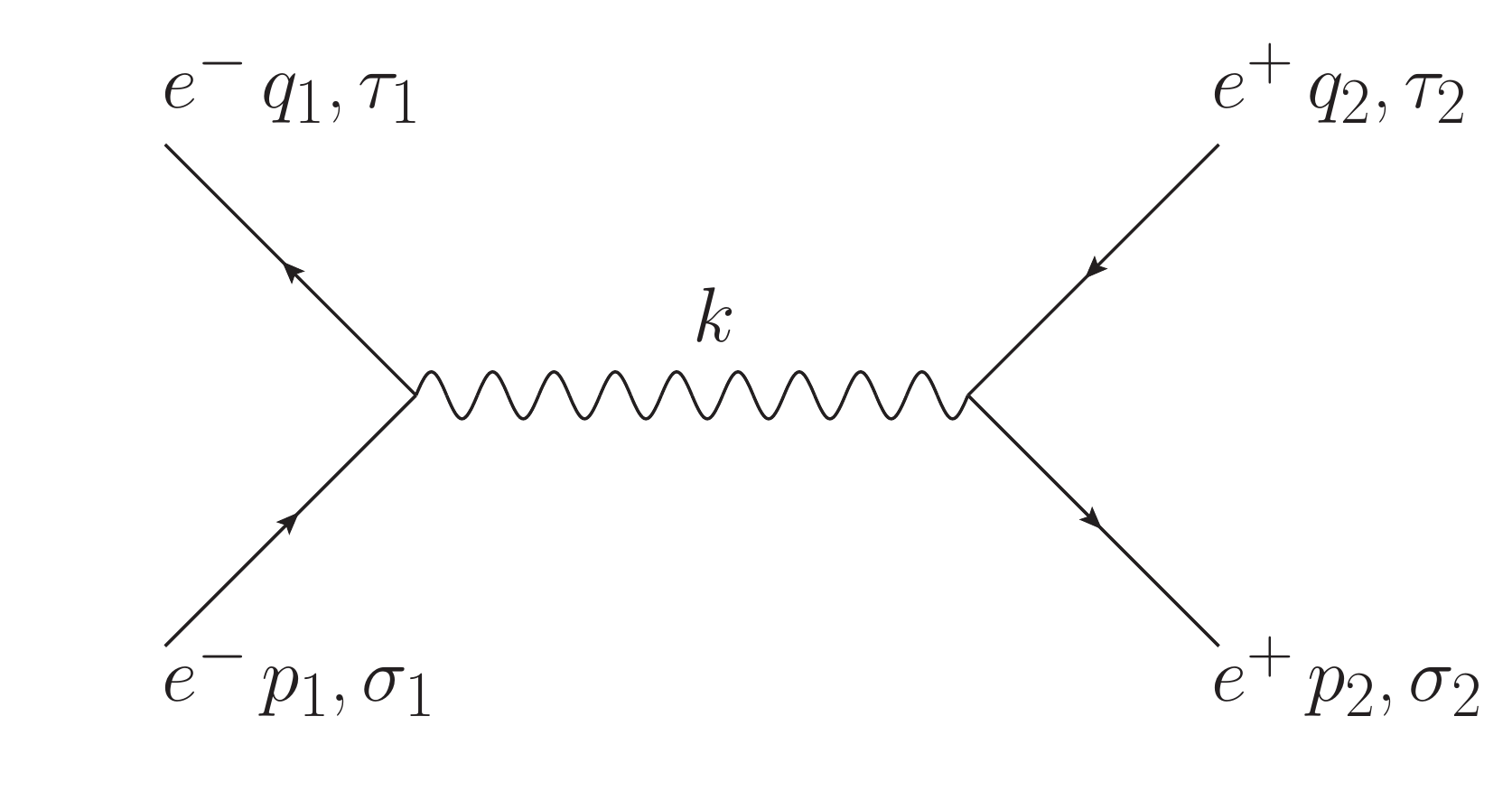}%
\hfill%
\includegraphics[width=0.35\columnwidth]{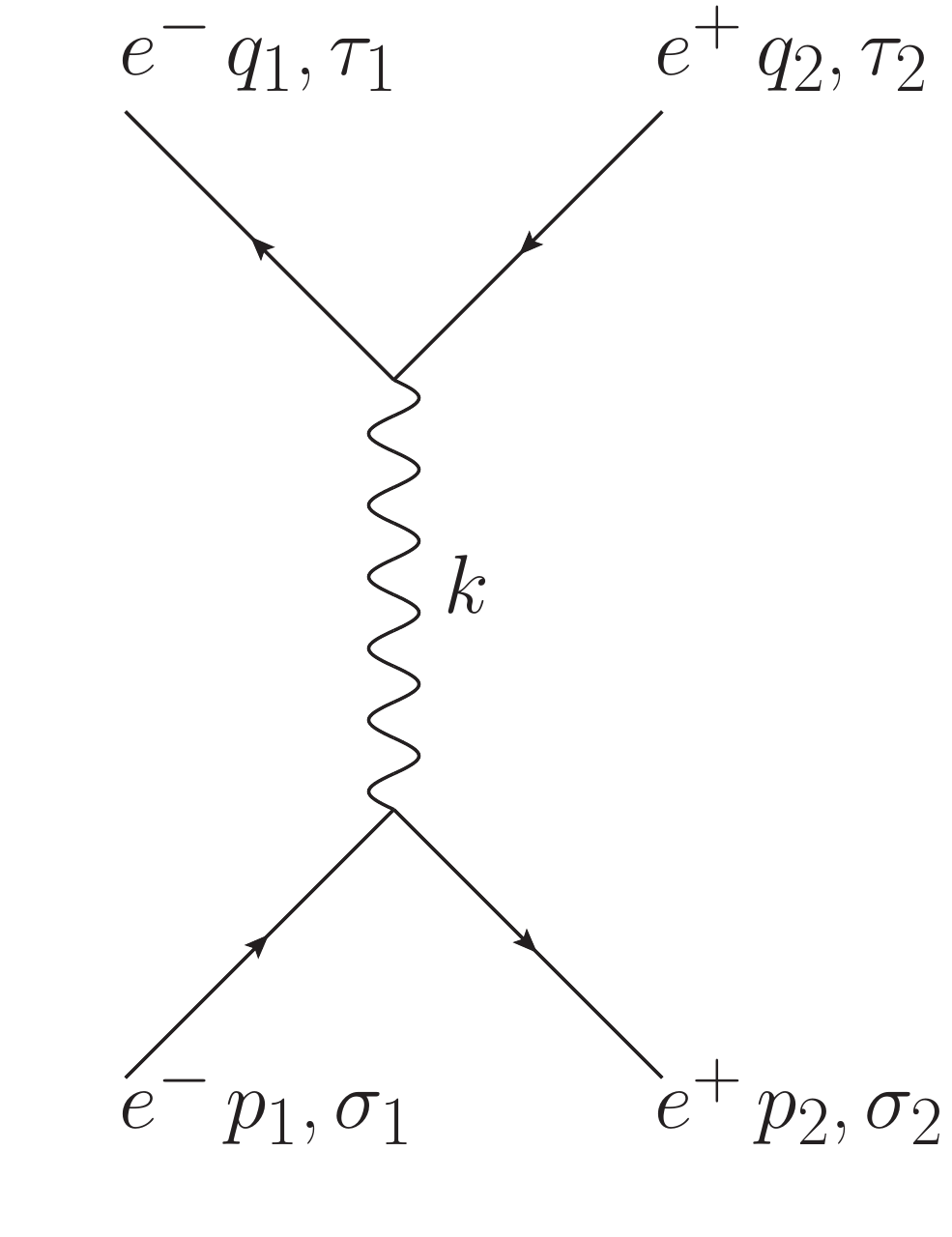}%
\caption{Feynman diagrams corresponding to the two terms in the $e^- e^+$ scattering amplitude. The 
left diagram corresponds to the $t$ channel (photon exchange) and the right 
diagram to the $s$ channel (pair annihilation).}%
\label{fig:feynman_graphs}%
\end{figure}

Let us first review how the Coulomb potential is obtained in the nonrelativistic
limit of QED from the application of Feynman rules to the electron-positron system. The 
scattering-matrix $S_{fi}$, from which the interaction potential may be obtained, is given by 
\begin{equation}
 \label{eq:S-matrix}
 S_{fi} \;\equiv\; \langle f | i \rangle \;=\; \delta_{fi} \,+\, 
i (2 \pi)^4 \, \delta^{(4)}(Q-P)\, T_{fi} \,,
\end{equation}
where $Q$ and $P$ correspond respectively to the final and initial total
momentum and $T_{fi}$ is the scattering amplitude. The two tree-level Feynman
diagrams contributing to $T_{fi}$ (see Fig.\ \ref{fig:feynman_graphs}) correspond to 
the $t$ and $s$ channels,
respectively coming from scattering with one photon exchange and to 
annihilation and creation of an $e^- e^+$ pair. We get
\begin{equation}
\label{eq:QED_Scattering_Amplitude}
 T_{fi} \;=\; \frac{1}{(2 \pi)^6} \frac{m^2}{\sqrt{E_{p_1}E_{p_2}E_{q_1}E_{q_2}}} 
 \left(t_{\text{exch}}+t_{\text{annihil}} \right)\,,
\end{equation}
where
\begin{align}
 t_{\text{exch}} \;=&\; e^2\,\overline{u}(q_1,\tau_1)\,\gamma^\mu 
  \,u(p_1,\sigma_1)\;P_{\mu \nu}(k)\; \nonumber \\
  \quad &  \times \,\overline{v}(p_2,\sigma_2)\,\gamma^\nu\,v(q_2,\tau_2) 
\end{align}
and
\begin{align}
 t_{\text{annihil}} \;={}&\;-e^2\,\overline{v}(p_2,\sigma_2)\,\gamma^\mu
  \,u(p_1,\sigma_1)\;P_{\mu \nu}(k)\; \nonumber \\
  \quad & \times \, \overline{u}  (q_1,\tau_1)\,\gamma^\nu v(q_2,\tau_2) \;.
\end{align}
We follow the notation in \cite{bjorken1965relativistic, Lucha:1991vn, Lucha:1995zv}:
$p_i$ denotes the momentum of the incoming particles and $q_i$ of the outgoing ones.
The particles' initial and final spins are respectively $\sigma_i$ and $\tau_i$.
We represent the photon propagator by a function $P_{\mu \nu}(k)$ of the photon momentum $k$.

We then make the nonrelativistic approximation, i.e.\ we impose the kinetic 
energy of the system to be much smaller than its rest energy 
($ |\vec{p}| \ll m \cong E$). The four-component state vectors become
\begin{align}
u(p,1/2) &\;\cong\;
	\begin{pmatrix}
		1 \\ 0 \\ 0 \\ 0
	\end{pmatrix}\;, \quad
	\quad
u(p,-1/2) \;\cong\;
	\begin{pmatrix}
		0 \\ 1 \\ 0 \\ 0
	\end{pmatrix}
\label{eq:spinors_non_relativistic1}
\end{align}
and
\begin{align}
v(p,1/2) 	&\;\cong\;
	\begin{pmatrix}
		0 \\ 0 \\ 0 \\ 1 
	\end{pmatrix}\;, \quad
	\quad
v(p,-1/2) 	\;\cong\;
	\begin{pmatrix}
		0 \\ 0 \\ -1 \\ 0
	\end{pmatrix}\,.
\label{eq:spinors_non_relativistic2}
\end{align}

In this approximation, $T_{fi}$ can be written as
\begin{equation}
 T_{fi} \;=\; \frac{1}{(2 \pi)^6}\left(t_\text{exch}+t_\text{annihil}\right)\,.
\end{equation}
To compute the exchange term, we adopt the Dirac representation for the gamma matrices and the 
center-of-momentum frame, obtaining
\begin{align}
t_{\text{exch}} \,&=\, e^2\,\delta^{\mu 0} \delta_{\sigma_1 \tau_1} \,P_{\mu \nu}(k) \,\delta^{\nu 0} \delta_{\sigma_2 \tau_2} \nonumber \\ 
		\,&=\, e^2\, P_{0 0}(k)\, \delta_{\sigma_1 \tau_1} \delta_{\sigma_2 \tau_2}\,,
\label{eq:texch}
\end{align}
with\footnote{We are using the metric $g_{\mu \nu} = \diag(-1,1,1,1)$, which will be 
more convenient when we consider Wick rotations later.}
\begin{equation}
k \,=\, p_1\, - \,q_1\, = \begin{pmatrix} 0 ,& \bf{k}\end{pmatrix}\,.
\label{eq:transferred_momentum}
\end{equation}
For the annihilation term, note that conservation of momentum at the vertices implies that
\begin{equation}
k \,=\, \begin{pmatrix} 2 m ,& 0\end{pmatrix}\,.
\label{eq:transferred_momentum_annih}
\end{equation}
For QED, the Feynman-gauge propagator is given by the expression $P_{\mu \nu}(k) = - g_{\mu \nu}/k^2$. 
As seen in Eqs.\ (\ref{eq:spinors_non_relativistic1}) and (\ref{eq:spinors_non_relativistic2}), the 
spinors are momentum-independent
in the nonrelativistic approximation and, while $t_{\rm exch}$ is proportional to
$1/\vec{k}^2$, we see that $t_{\rm annihil}$ will be proportional to $1/4 m^2$.
Thus, we can neglect annihilation effects and the scattering amplitude is given by
\begin{equation}
T_{fi} \;=\; \frac{1}{(2 \pi)^6}\frac{e^2}{\vec{k}^2}\,.
\label{eq:scatter_amp}
\end{equation}
 
The potential can then be obtained as an inverse Fourier transform, which leads to the 
Coulomb potential
\begin{align}
 V(\vec{r}) 	\,=\, - (2 \pi)^3 \int \exp(-i \vec{k}\cdot\vec{r})\;T_{fi}(\vec{k}^2)\,d^3 k \,=\,- \frac{e^2}{4 \pi 	r}\,.
\label{eq:Couloumb_Pot}
\end{align}

\vskip 3mm
For QCD, we replace the photon by the gluon and the electron-positron pair by a quark-antiquark pair.
The scattering amplitude will continue to be expressed as a sum of the two terms, now given by
\begin{align}
 \label{eq:QCD_Scattering_Amplitude}
t_{\text{exch}} \,&=\,\phantom{\,+\,}g_0^2\;\overline{u}(q_1,\tau_1)\,c_{1,\,f}^\dagger\,\lambda^a\gamma^\mu \,c_{1,\,i}\,u(p_1,\sigma_1)\;P_{\mu \nu}^{a b}(k)\; \nonumber \\[1mm]
		              &\phantom{ = \,+\,}\times\, \overline{v}(p_2,\sigma_2)\,c_{2,\,i}^\dagger\,\lambda^b\gamma^\nu\,c_{2,\,f}\,v(q_2,\tau_2)
\end{align}
and
\begin{align}
t_{\text{annihil}} \,&=\,-\,g_0^2\;\overline{v}(p_2,\sigma_2)\,c_{2,\,f}^\dagger\,\lambda^a\gamma^\mu  \,c_{1,\,i}\,u(p_1,\sigma_1)\;P_{\mu \nu}^{a b}(k)\; \nonumber \\[1mm]
	      &\phantom{=\,-\,}\times\,\overline{u}(q_1,\tau_1)\,c_{1,\,f}^\dagger\,\lambda^b\gamma^\nu c_{2,\,f}\,v(q_2,\tau_2)\,,
\end{align}
where $c_{(1,2),(i,f)}$ are three-component color vectors and $\lambda^a$ are
the Gell-Mann matrices. 

With respect to Eq.\ (\ref{eq:QED_Scattering_Amplitude}), the terms $t_{\rm exch}$ and $t_{\text{annihil}}$
will have multiplicative (Casimir) factors, coming from the sum over colors. This sum is obtained assuming that the 
incoming/outgoing quarks and antiquarks have equal probability of being in a given color state and
imposing a color-diagonal gluon propagator. The factors are given respectively by
\begin{align}
\label{eq:color_factor_exch}
 c^\dagger_{1,f}\,\lambda^a\,c_{1,i}\;c^\dagger_{2,i}\,\lambda^a\,c_{2,f} \;=\; 
 \frac{1}{3} \Tr \lambda^a \lambda^a \;=\; \frac{\delta^{a a}}{6} \;=\; \frac{4}{3}\,
\end{align}
and
\begin{align}
\label{eq:color_factor_annih}
c^\dagger_{2,i}\,\lambda^a\,c_{1,i}\;c^\dagger_{1,f}\,\lambda^a\,c_{2,f} \;=\;
 \frac{1}{3} (\Tr \lambda^a) (\Tr \lambda^a) \;=\; 0\,.
\end{align}

Therefore, annihilation effects do not contribute, independently of the nonrelativistic approximation.
If we now assume a free (i.e.\ tree-level) gluon propagator
\begin{equation}
P_{\mu \nu}^{a b} \;=\; -\frac{g_{\mu \nu}\, \delta^{a b}}{k^2}\,, 
\end{equation}
we obtain a Coulomb-like interquark potential
\begin{equation}
V(r) \;=\; -\frac{4}{3}\frac{g_0^2}{4 \pi r} \;=\; -\frac{4}{3}\frac{\alpha_s}{r}\,.
\label{eq:QCD_pert_pot}
\end{equation}

Notice that the above potential is non-confining.
This is expected, since it results from a perturbative calculation, while confinement 
is a nonperturbative phenomenon. The addition of a linear term as described in Section \ref{sec:Introduction}
leads to the Cornell, or Coulomb-plus-linear, potential \cite{Eichten:1978tg,Eichten:1979ms,Eichten:2002qv}
\begin{equation}
V(r) \;=\; -\frac{4}{3}\frac{\alpha_s}{r} \,+\, \sigma r\,,
\label{eq:cornell_pot}
\end{equation}
which describes surprisingly well the states of charmonium and bottomonium.

\vskip 3mm
As mentioned in Section \ref{sec:Introduction}, we substitute the free propagator by a 
fully nonperturbative one in the OGE term. More precisely, we use the propagator 
\begin{align}
\label{eq:lattice_gluon_propagator}
 P_{\mu \nu}^{a b}(k) \;=\; \frac{C\,(s+k^2)}{t^2+u^2 k^2+k^4} \left(\delta_{\mu \nu} - \frac{k_\mu k_\nu}{k^2}\right) \delta^{a b}\,, \\[2mm]
  \begin{aligned}
   C & \;=\; \num{0.784},\; & s &  \;=\; \SI{2.508}{\giga \electronvolt \squared}\,, \nonumber \\
   u & \;=\; \SI{0.768}{\giga \electronvolt},\; & t & \;=\; \SI{0.720}{\giga \electronvolt \squared}\,,
  \end{aligned}
\end{align}
obtained from fits of 
lattice data for a pure $SU(2)$ gauge theory in Landau gauge given in Ref.\ \cite{Cucchieri:2011ig}.
We note that the lattice data for propagators in the $SU(2)$ and $SU(3)$ case have essentially the same 
behavior apart from a global constant \cite{Cucchieri:2007zm}. Also, the above parameters correspond to a
value $1/k^2$ at 2 \si{\giga \electronvolt}. Here we choose to normalize the propagator to $1/k^2$ at $k\to\infty$, i.e.\ we
adopt $C=1$.

We now follow the same procedure as in the QED case. From Eq.\ (\ref{eq:texch}) we notice that, in
the nonrelativistic approximation, only the component $P_{0 0}(0,\vec{k})$ survives in the $t_{\text{exch}}$ term and thus
the term $k_\mu k_\nu/k^2$ vanishes [see Eq.\ (\ref{eq:transferred_momentum})]. Lastly, in order to convert the propagator in 
Eq.\ (\ref{eq:lattice_gluon_propagator}), which was obtained in Euclidean space,
to Minkowski space, we undo the Wick rotation, taking $\delta_{\mu \nu} \rightarrow -g_{\mu \nu}$. We obtain\footnote{Let 
us recall that the propagator is a gauge-dependent quantity. A gauge-independent
potential obtained from the (Coulomb-gauge) propagator is discussed in \cite{Popovici:2010fy}.}
\begin{equation}
 P_{0 0}^{a b}\big(\vec{k}\,\big) \;=\; \frac{C\,\big(s+\vec{k}^2\big)}{t^2+u^2 \vec{k}^2+\vec{k}^4}\, 
 \delta^{a b}\,.
\label{eq:Nonrelativistic_Lattice_Propagator}
\end{equation}
This leads us to the following scattering amplitude
\begin{equation}
T_{fi} \;=\; \frac{4}{3}\,\frac{g_0^2}{(2 \pi)^6}\,\frac{C\,\big(s+\vec{k}^2\big)}{t^2+u^2 \vec{k}^2+\vec{k}^4}\,.
\label{eq:scatter_amp_QCD}
\end{equation}

The potential is obtained, as was done in the QED case, as a Fourier transform of the scattering amplitude
[see Eq.\ (\ref{eq:Couloumb_Pot})]. We use spherical coordinates for $\vec{k}$ and set $\vec{r} = r\, \hat{z}$.
The angular integration is then trivial, resulting in\footnote{For the evaluation of this integral only, we
will denote $\abs{\vec{k}}=k$.}
\begin{align}
V(\vec{r}\,) \,=\, -\frac{4}{3}\,\frac{g_0^2 C}{2i(2 \pi)^2 r} 
           \int_{-\infty}^\infty{\frac{\left(s+k^2\right) \left(e^{i k r}-e^{-i k r}\right)}{t^2+u^2 k^2+k^4} k\,dk}\,.
\label{eq:angular_integration}
\end{align}
The integral in Eq.\ (\ref{eq:angular_integration}) can be solved using residue calculations. The four poles 
in the integrand are symmetrically distributed in the four quadrants of the complex
plane. We index these poles in the following way
\begin{align}
k_{m,n} \;&=\; (-1)^m\, i \sqrt{t}\, \exp{\left[(-1)^n\,i \frac{\theta}{2}\right]}\,,\quad m, \,n = 0,1\,,
\end{align}
where
\begin{align}
 \theta \;&\equiv\; \arctan \left(\frac{\sqrt{4 t^2-u^4}}{u^2}\right)\,.
\end{align}
The associated contour integral is performed by considering its two terms separately:
for the term with $e^{i k r}$ (respectively with $e^{-i k r}$) we close the contour 
above (respectively below). The residues are given by
\begin{align}
 &\Res \left[\frac{\left(s+k^2\right)k\,e^{\pm i k r}}{t^2+u^2 k^2+k^4},\, k_{m,n}\right] 
 \;=\; \frac{1}{2}\,\frac{\left(s+k_{m,n}^2\right)\,e^{\pm i k_{m,n} r}}{ u^2+2 k_{m,n}^2}\,.
\end{align}
The result is simplified by noticing that 
$\,k_{1,0} = -k_{0,0}$, $\,k_{0,1}=-k_{1,1}$ and $\,k_{1,1}=k_{0,0}^*$. 
The obtained potential, which we call the lattice-gluon-propagator potential $V_{LGP}$, is given by 
\begin{equation}
V_{LGP}(r) \;=\; -\frac{4}{3} \frac{2\alpha_s}{r}
 \Re \left[ \frac{C(s+k_{0,0}^2)\,e^{i k_{0,0} r}}{u^2+2 k_{0,0}^2}\right]\,,
\label{eq:Potential_QCD}
\end{equation}
where $\alpha_s = g_0^2/4 \pi$ [see Eq.\ (\ref{eq:QCD_pert_pot})].
\vskip 3mm

\begin{figure}[t]
\centering
\caption{Comparison between the lattice-gluon-propagator potential $V_{\mbox{LGP}}$ and the Coulomb-like 
potential (color factor included) in the charmonium case.}
\label{fig:Comparison_Potentials}
\vspace{2mm}
 \includegraphics{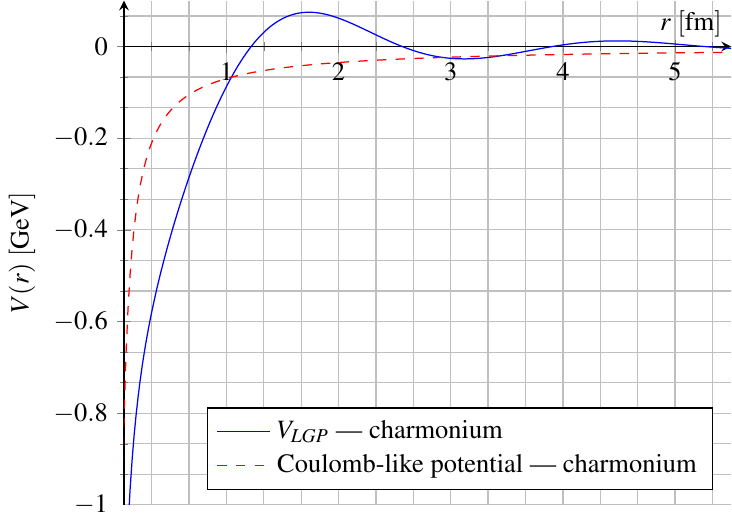}
\end{figure}

In order to use the above expression in our spectrum calculation, we use the four-loop formula 
available in Ref.\ \cite[Section 9]{Agashe:2014kda} as well as the $\Lambda_{\mbox{QCD}}$
values (also available in Ref.\ \cite[Section 9]{Agashe:2014kda}) to evaluate the strong coupling constant
$\alpha_s$ at the energy scale of 
the mass of the 1S quarkonia states [respectively $J/\psi$ and $\Upsilon$(1S) 
in the charmonium and bottomonium cases].
The resulting potential is compared (for the charmonium case) to the Coulomb-like potential from
Eq.\ (\ref{eq:QCD_pert_pot}) in Fig.\ \ref{fig:Comparison_Potentials}.
We see that, although the two curves are clearly different, $V_{LGP}$ is also non-confining. 
Thus, since (tree-level) perturbation theory was applied to obtain this potential,
the property of confinement was lost, even though the used propagator was obtained nonperturbatively. 

We therefore add a linearly rising term 
$\sigma r$ to the potential in order to model confinement, as done for the Cornell-potential case. 
The resulting expression is the lattice-gluon-propagator-plus-linear potential
\begin{equation}
V_{LGP+L}(r)\;\equiv\;V_{LGP}(r)\,+\,\sigma r\,.
\label{eq:lgp+l_potential}
\end{equation}
Also, the nonrelativistic approximation removes any spin dependence from the interactions. This means that, in 
our description, states with different spin values will be degenerate.

\section{Method for Obtaining Quarkonium Masses}
\label{sec:Methods}
 
Consider a central (nonrelativistic) potential describing the interaction between two particles. 
Since we are dealing with a two-particle system, we can write 
the Hamiltonian in terms of relative coordinates and use separation of variables 
in the resulting partial differential equation to isolate the angular part 
of the wave function, given by the spherical harmonics. Lastly, 
we perform the usual substitution of variables in the radial wave function
$R(r) = f(r)/r$ to obtain the ordinary differential equation (ODE) for $f(r)$
\begin{equation}
\label{eq:Radial_Equation}
\frac{d^2 f}{dr^2}\,+\,2\mu\left[E - V(r) -2m - \frac {l\left(l+1\right)}{2 \mu r^2} \right] f\left(r\right) \;=\; 0 \,,
\end{equation}
where $\mu$ is the reduced mass
\begin{equation}
\mu \;=\; \frac{m}{2}\,,
\end{equation}
and $m$ is the mass of the heavy (charm or bottom) quark.
We use units such that $c = \hbar = 1$.
Notice as well the addition of the rest mass of the particles, which will allow us to 
compare the eigenvalue directly with the mass values given in Ref.\ \cite{Agashe:2014kda}.

The above ODE has to be solved with proper boundary-value conditions. The first one is that
$f(0) = 0$. This comes from the requirement that $R(0)$ be non-singular. A second
condition is that $f(r \to \infty) = 0$ and comes from the fact that $R(r)$ is
normalized, i.e.
\begin{equation}
\int_0^\infty \left| R(r) \right|^2 r^2 dr \;=\; \int_0^\infty \left| f(r) \right|^2 dr = 1\,.
\end{equation}

In the limit of large $r$, the potential is
dominated by the linearly rising term and the ODE becomes
\begin{equation}
	\frac{d^2 f(r)}{dr^2}\,-\,2\mu \sigma r f\left(r\right) \;=\; 0 \,.
	\label{eq:ODE_large_r}
\end{equation}

The general solution of this equation is the linear combination of the Airy functions $Ai(\rho)$ and 
$Bi(\rho)$ \cite{abramowitz1964handbook}, where
$\,\rho = (2 \mu \sigma)^{1/3} \,r$. However, the Airy function of the second kind $\Bi(\rho)$ diverges at 
large $\rho$ and therefore it does not obey the boundary condition at infinity. 
For $\rho>0$, the Airy function of the first kind can be written as
\begin{equation}
	\Ai(\rho) = \frac{1}{\pi}\sqrt{\frac{\rho}{3}} \BesselK_{1/3}\left(\frac{2}{3}\rho^{3/2}\right)\,,
	\label{eq:Airy_Function}
\end{equation}
where $\BesselK_\nu(x)$ is the modified Bessel function of the second kind. One can try the Ansatz 
$f(\rho) = g(\rho) \Ai(\rho)$ and use the property $\BesselK_\nu'(x) = \nu \BesselK_\nu(x)/x-K_{\nu+1}$
in the ODE in Eq.\ (\ref{eq:Radial_Equation}) to obtain a second-order ODE with coefficients in terms of 
$\Ai(\rho)$ and $\BesselK_{4/3}(x)$. However, by expressing these functions as a power series in $\rho$,
one clearly sees that an analytic solution would be challenging, even for 
the simpler case of the Cornell potential. We therefore seek a numerical solution of the problem.

To this end, we use the so-called shooting method \cite{press2007numerical}. It consists in picking trial values in a discretized
range for the eigenenergies, integrating the ODE for each of these values to obtain the corresponding wave 
function and choosing the energies for which the wave function obeys the boundary conditions approximately. 
We use the backward second-order Runge-Kutta method to integrate the wave
function, starting from a maximum value $r_{max}$ for the radial coordinate until the origin, in steps 
of $dr$ (we adapt the method as presented in Ref.\ \cite{press2007numerical} 
by adopting a negative integration step). We choose $r_{max}$ sufficiently 
large so that we can use $f(r_{max}) = \Ai(r_{max})$ and $f'(r_{max}) = \Ai'(r_{max})$ as initial conditions. 
In practice, the wave 
function will not obey the boundary conditions exactly since the proposed energy is unlikely to be an exact eigenenergy. 
Nevertheless, we may count the number of nodes of the wave function: each time we observe an increase in the 
number of nodes when compared with the previously proposed energy, the desired eigenenergy will be between the 
two proposed values. 
We further refine our method by adapting the bisection method to search for the eigenenergy in this interval, 
thus allowing the use of a coarse grid without loss of precision.

We test our algorithm with the parameters of the Cornell potential set to 
$\sigma = \SI{1}{\giga \electronvolt \squared}$, $2\mu = \SI{1}{\giga \electronvolt} $ and $4 \alpha_s/3 = 1$. 
These parameters are the ones used in Ref.\ \cite{2014arXiv1411.2023H}, 
which adopts a different approach for solving the problem. We find agreement with their values
up to the 4th and in some cases even 5th decimal place. Similarly, Ref.\ \cite{Chung:2008sm} uses yet another 
numerical method to compute the 
eigenenergies for a different set of parameters, allowing comparison with our results.\footnote{More precisely,
with respect to our Tables \ref{tb:results_simultaneous_fit_charm}, \ref{tb:results_simultaneous_fit_bottom} 
and \ref{tb:parameters_results_simultaneous_fit} in Section \ref{sec:Results}, we find agreement up to the 
3rd decimal places. We must consider that, in this comparison, our parameters are close to but not identical
to the ones used in Ref.\ \cite{Chung:2008sm},
which might explain the slightly worse agreement than in the comparison with \cite{2014arXiv1411.2023H}.}

\vskip 3mm
We consider as free parameters in the ODE the string tension $\sigma$
and the mass of the quark. The reason for including the quark mass as a parameter
is that quark masses are not observable directly and depend on the renormalization scheme.
Therefore, we have the freedom to pick one that best describes the observed spectrum.

To find the best values for these parameters, we set up a two-dimensional grid of
values of $m$ and $\sigma$. Then, we compute the eigenenergies for each proposed 
set of parameters and select the one that best describes the observed spectrum.
As a criterion for choosing the optimal parameters we consider the minimization of 
$\chi^2$ in the description of a few input values from experiment, i.e.\
we pick the set of parameters minimizing
\begin{equation}
	\chi^2(\text{parameters}) \;=\; \sum_{i} \left(\frac{E_{i}-E_{i,\text{experimental}}}{\sigma_i}\right)^2\,,
	\label{eq:residual_criteria}
\end{equation}
where $\sigma_i$ is the experimental error associated with the energy $E_{i,\text{experimental}}$ and the $E_i$'s
are the eigenenergies computed as described above.

Notice that the numerical method described above can be applied to any central potential. 
Therefore, we can obtain the results for the 
Cornell and the $V_{LGP+L}$ potential using the same algorithm, with the same inputs, making it easy to 
compare the differences between the two cases.
 
\section{Results}
\label{sec:Results}

We apply our method to two closely nonrelativistic systems: charmonium and bottomonium.
Since the bottomonium is heavier in comparison with its kinetic energy, we expect that the proposed model 
will work better for it than for charmonium.

As stated at the end of Section \ref{sec:Pot_Model_Review}, this model does not see spin interactions. This 
results in states with high degeneracy, in comparison with the experimental data. In order to perform a fit to tune
the parameters of our model, we need to average over these states with different spin. This is done in 
Ref.\ \cite{Olsson:1994cv} by using the degeneracy of each state as a weight, i.e.\ the spin-averaged mass
of the states with principal quantum number $n$ and in the $X$-wave state ($X = S,P,D \dots$) is given by 
\begin{equation}
\langle M(nX) \rangle \; = \; \frac{\sum_{i=1}^{N_l} m_i(nX) g_i}{\sum_{i=1}^{N_l} g_i}\,,
\label{eq:spin_average}
\end{equation}
where $m_i(nX)$ is the mass of each of the $N_l$ states with the same $l$ and
$g_i$ is the degeneracy of the state.
The uncertainty associated with the above average may be estimated by propagation of errors, where the error
of each mass is due to the width of the resonance peak\footnote{We recall that bound states are detected 
by plotting a histogram of number of particles (cross-section) detected in a collision versus the energy of 
the collision. When a resonance is found, it is associated to a bound state.}. This leads to very small
errors, and we find that this is not a good way to estimate our uncertainties, as discussed below.
We refer to this averaging procedure as ``Mass Input 1''.

Instead, we choose to average the spins by imagining that, if
the experiments were not very precise, we would not see several narrow nondegenerate states, but 
broad degenerate ones, i.e.\ a low precision experiment would see the peaks merged.
We thus consider as input for a state the midpoint between the state with lowest energy and the 
one with highest energy. The error is estimated as half of the distance between these two states.
We refer to this method as ``Mass Input 2''.

In Tables \ref{tb:input_data_charm} and \ref{tb:input_data_bottom} we show the data extracted from 
Ref.\ \cite{Agashe:2014kda} for charmonium and bottomonium, combining the different
spin states using the two methods described above (i.e.\ ``Mass Input 1'' and ``Mass Input 2'').
We choose to include just the states present in the meson summary table of Ref.\ \cite{Agashe:2014kda}
that are regarded as established particles. We omit charged states from our table, since quarkonia states 
must be neutral.
We remark that, for our fit, we use only the states $1S$, $1P$ and $2S$ of charmonium and bottomonium.

\begin{table*}[h!]%
\centering
\caption{Experimental spectrum of charmonium and input values considered in our fits (see text).}
\label{tb:input_data_charm}
\begin{tabular}{cS[table-format=2.9]ccS[table-format=2.9]S[table-format=+2.9]}
\hline
Particle Name 		& {Mass (\si{\giga \electronvolt})}	& $J^{P C}$	& $l$ 			& {Mass Input 1(\si{\giga \electronvolt})}				& {Mass Input 2(\si{\giga \electronvolt})}			\\ \hline
$\eta_c(1S)$		& \EtacOneS +- \EtacOneSErr		& $0^{- +}$	& \multirow{2}{*}{0}	& \multirow{2}{*}{\tablenum{\InpCharmOneS +- \InpCharmOneSErr}}	& \multirow{2}{*}{\tablenum{3.040 +- .057}}\\
$J/\psi(1S)$		& \PsiOneS +- \PsiOneSErr		& $1^{- -}$	& 			& 									& \\ \hline
$\chi_{c0}(1P)$ 	& \chicZOneP +- \chicZOnePErr 		& $0^{+ +}$	& \multirow{4}{*}{1}	& \multirow{4}{*}{\tablenum{\InpCharmOneP +- \InpCharmOnePErr}}	& \multirow{4}{*}{\tablenum{3.485  +- .070}}\\
$\chi_{c1}(1P)$		& \chicOOneP +- \chicOOnePErr		& $1^{+ +}$	& 			&									& \\
$h_c(1P)$		& \hcOneP +- \hcOnePErr			& $1^{+ -}$	& 			&									& \\
$\chi_{c2}(1P)$		& \chicTOneP +- \chicTOnePErr		& $2^{+ +}$	& 			&									& \\ \hline
$\eta_c(2S)$		& \EtacTwoS +- \EtacTwoSErr		& $0^{- +}$	& \multirow{2}{*}{0}	& \multirow{2}{*}{\tablenum{\InpCharmTwoS +- \InpCharmTwoSErr}}	& \multirow{2}{*}{\tablenum{ 3.663 +- .023}}\\
$\psi(2S)$		& \PsiTwoS +- \PsiTwoSErr		& $1^{- -}$	& 			& 									& \\ \hline
$\psi(3770)$		& \PsiA +- \PsiAErr			& $1^{- -}$	&	0 or 2		& \multirow{1}{*}{\tablenum{\PsiA +- \PsiAErr}}			& \multirow{1}{*}{\tablenum{\PsiA +- \PsiAErr}} \\ \hline
$X(3872)$		& \XA +- \XAErr				& $1^{+ +}$	& 1 			& \multirow{1}{*}{\tablenum{\XA +- \XAErr}}				& \multirow{1}{*}{\tablenum{\XA +- \XAErr}} \\ \hline
$\chi_{c0}(2P)$		& \chicZTwoP +- \chicZTwoPErr		& $0^{+ +}$	& \multirow{2}{*}{1}	& \multirow{2}{*}{\tablenum{3.9257 +- 0.0025}}				& \multirow{2}{*}{\tablenum{ 3.9228 +-0.0044}}\\ 
$\chi_{c2}(2P)$		& \chicTTwoP +- \chicTTwoPErr		& $2^{+ +}$	& 			& 									& \\ \hline
$\psi(4040)$		& \PsiB +- \PsiBErr			& $1^{- -}$	& 0 or 2		& \multirow{1}{*}{\tablenum{\PsiB +- \PsiBErr}}			& \multirow{1}{*}{\tablenum{\PsiB +- \PsiBErr}}\\ \hline
$\psi(4160)$		& \PsiC +- \PsiCErr			& $1^{- -}$	& 0 or 2		& \multirow{1}{*}{\tablenum{\PsiC +- \PsiCErr}}			& \multirow{1}{*}{\tablenum{\PsiC +- \PsiCErr}}	\\ \hline
$X(4260)$ 		& \XB +- \XBErr				& $1^{- -}$	& 0 or 2		& \multirow{1}{*}{\tablenum{\XB +- \XBErr}}				& \multirow{1}{*}{\tablenum{\XB	+- \XBErr}}\\ \hline
$X(4360)$ 		& \XC +- \XCErr				& $1^{- -}$	& 0 or 2		& \multirow{1}{*}{\tablenum{\XC +- \XCErr}}				& \multirow{1}{*}{\tablenum{\XC +- \XCErr}}\\ \hline
$\psi(4415)$		& \PsiD +- \PsiDErr			& $1^{- -}$	& 0 or 2		& \multirow{1}{*}{\tablenum{\PsiD +- \PsiDErr}}			& \multirow{1}{*}{\tablenum{\PsiD +- \PsiDErr}}\\ \hline
$X(4660)$		& \XD +- \XDErr				& $1^{- -}$	& 0 or 2		& \multirow{1}{*}{\tablenum{\XD +- \XDErr}}				& \multirow{1}{*}{\tablenum{\XD	+- \XDErr}}\\ \hline
\end{tabular}
\end{table*}
\begin{table*}[h!]%
\centering
\caption{Experimental spectrum of bottomonium and input values considered in our fits (see text). 
We include the unconfirmed state $\eta_b(1S)$ since we need
it to improve our results, as discussed in the text.}
\label{tb:input_data_bottom}
\begin{tabular}{cS[table-format=2.9]ccS[table-format=2.9]S[table-format=+2.9]}
\hline
Particle Name 		& {Mass (\si{\giga \electronvolt})}	& $J^{P C}$	& $l$ 			& {Mass Input 1(\si{\giga \electronvolt})}				& {Mass Input 2(\si{\giga \electronvolt})}			\\ \hline
$\eta_b(1S)$		& \EtabOneS +- \EtabOneSErr		& $0^{- +}$	& \multirow{2}{*}{0}	& \multirow{2}{*}{\tablenum{9.4447 +- 0.0010}}				& \multirow{2}{*}{\tablenum{9.429 +- .031}}\\ 
$\Upsilon(1S)$		& \UpsilonOneS +- \UpsilonOneSErr	& $1^{- -}$	& 			&									& \\ \hline
$\chi_{b0}(1P)$		& \chibZOneP +- \chibZOnePErr		& $0^{+ +}$	& \multirow{4}{*}{1}	& \multirow{4}{*}{\tablenum{\InpBottomOneP +- \InpBottomOnePErr}}	& \multirow{4}{*}{\tablenum{ 9.886 +- .026}}\\
$\chi_{b1}(1P)$		& \chibOOneP +- \chibOOnePErr		& $1^{+ +}$	&			&									& \\
$h_b(1P)$		& \hbOneP +- \hbOnePErr			& $1^{+ -}$	&			&									& \\
$\chi_{b2}(1P)$		& \chibTOneP +- \chibTOnePErr  	& $2^{+ +}$	&			&									& \\ \hline
$\Upsilon(2S)$		& \UpsilonTwoS +- \UpsilonTwoSErr	& $1^{- -}$	&	0		& \multirow{1}{*}{\tablenum{\UpsilonTwoS +- \UpsilonTwoSErr}}		& \multirow{1}{*}{\tablenum{\UpsilonTwoS +- \UpsilonTwoSErr}}\\ \hline
$\Upsilon(1D)$		& \UpsilonOneD +- \UpsilonOneDErr	& $2^{- -}$	&	2		& \multirow{1}{*}{\tablenum{\UpsilonOneD +- \UpsilonOneDErr}}		& \multirow{1}{*}{\tablenum{\UpsilonOneD +- \UpsilonOneDErr}}\\ \hline
$\chi_{b0}(2P)$		& \chibZTwoP +- \chibZTwoPErr		& $0^{+ +}$	& \multirow{3}{*}{1}	& \multirow{3}{*}{\tablenum{\InpBottomTwoP +- \InpBottomTwoPErr}}	& \multirow{3}{*}{\tablenum{ 10.251 +- .018}}\\
$\chi_{b1}(2P)$		& \chibOTwoP +- \chibOTwoPErr		& $1^{+ +}$	&			& 									& \\
$\chi_{b2}(2P)$		& \chibTTwoP +- \chibTTwoPErr		& $2^{+ +}$	&			& 									& \\ \hline
$\Upsilon(3S)$		& \UpsilonThreeS +- \UpsilonThreeSErr	& $1^{- -}$	&	0		& \multirow{1}{*}{\tablenum{\UpsilonThreeS +- \UpsilonThreeSErr}}	& \multirow{1}{*}{\tablenum{\UpsilonThreeS +- \UpsilonThreeSErr}}\\ \hline
$\chi_{b}(3P)$		& \chibThreeP +- \chibThreePErr	& $?^{? +}$	& 1			& \multirow{1}{*}{\tablenum{\chibThreeP +- \chibThreePErr}}		& \multirow{1}{*}{\tablenum{\chibThreeP +- \chibThreePErr}}\\ \hline	
$\Upsilon(4S)$		& \UpsilonFourS +- \UpsilonFourSErr	& $1^{- -}$	&	0		& \multirow{1}{*}{\tablenum{\UpsilonFourS +- \UpsilonFourSErr}}	& \multirow{1}{*}{\tablenum{\UpsilonFourS +- \UpsilonFourSErr}}\\ \hline
$\Upsilon(10860)$	& \UpsilonA +- \UpsilonAErr		& $1^{- -}$	& 0 or 2		& \multirow{1}{*}{\tablenum{\UpsilonA +- \UpsilonAErr}}		& \multirow{1}{*}{\tablenum{\UpsilonA +- \UpsilonAErr}}\\ \hline
$\Upsilon(11020)$	& \UpsilonB +- \UpsilonBErr		& $1^{- -}$	& 0 or 2		& \multirow{1}{*}{\tablenum{\UpsilonB +- \UpsilonBErr}}		& \multirow{1}{*}{\tablenum{\UpsilonB +- \UpsilonBErr}}\\ \hline
\end{tabular}
\end{table*}

We implement the algorithm described in Section \ref{sec:Methods} using the following parameters: 
for the {\bf charmonium} case, the quark mass ranges from \SI{1.0}{\giga \electronvolt} to 
\SI{2.0}{\giga \electronvolt} in steps of $\SI{0.01}{\giga \electronvolt}$,
the string-tension parameter $\sigma$ ranges from \SI{0.1}{\giga \electronvolt \squared} to 
\SI{0.5}{\giga \electronvolt \squared} in steps of $\SI{0.01}{\giga \electronvolt \squared}$ and 
we use\footnote{We actually compute and use $\alpha_s$ with 
20 decimal places to ensure precision, but we are aware that the error of the computation must be far greater.} 
$\alpha_s = 0.2663$.
The wave function is integrated from a maximum distance of $\SI{20.0}{\per \giga \electronvolt}$ until the 
origin in steps of $\SI{5.E-3}{\per \giga \electronvolt}$ and 
the eigenenergies are searched in the range from \SI{2.0}{\giga \electronvolt} to \SI{6.0}{\giga \electronvolt} 
in steps of \SI{0.04}{\giga \electronvolt}.
For the {\bf bottomonium} case, the quark mass ranges from \SI{4.15}{\giga \electronvolt} to 
\SI{4.85}{\giga \electronvolt} in steps of $\SI{0.01}{\giga \electronvolt}$, 
we use the same range as above for $\sigma$ and 
$\alpha_s = \num{0.1843}$. 
The parameters of integration for the wave function are the same as for the charmonium case and 
the eigenenergies are searched in the range from \SI{8.5}{\giga \electronvolt} to \SI{12.5}{\giga \electronvolt} 
in steps of \SI{0.04}{\giga \electronvolt}. 

\begin{figure}
\centering
\caption{Comparison of the Cornell Potential with $V_{LGP+L}$. For the value of the strong coupling
constant, we choose the one used in the (constrained-fit) description of the charmonium spectrum.}
 \label{fig:full_pot_comparison}
	\vskip 1mm
	\includegraphics{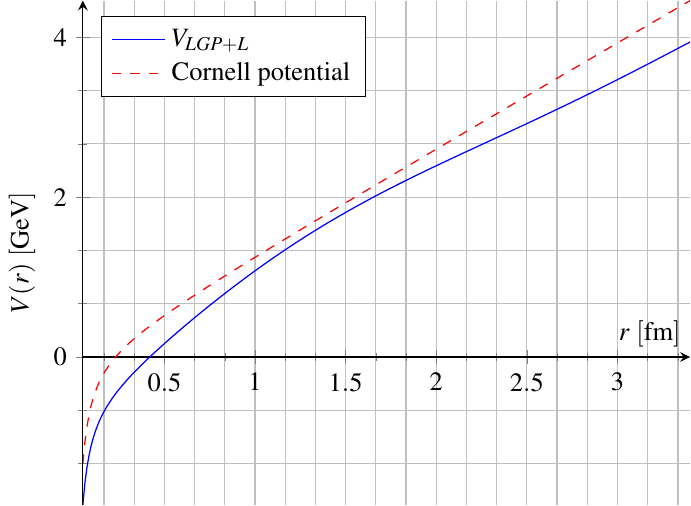}
\end{figure}

We first made fits for the charmonium and bottomonium spectra independently, using ``Mass Input 1''. 
This yields a very large value of $\chi^2$ for both systems (varying from 
$1.9 \times 10^2$/dof --- when we fit bottomonium with $V_{LGP+L}$ --- 
to $23 \times 10^3$/dof --- in the case where we fit
charmonium using Cornell potential). 
We thus conclude that the errors shown in the column ``Mass Input 1'' of Tables
\ref{tb:input_data_charm} and \ref{tb:input_data_bottom} are inconsistent with the overall precision of our calculation, considering e.g.\ the 
effects of the nonrelativistic approximation employed.
Instead, using ``Mass Input 2'' gives acceptable
values for $\chi^2$ in the charmonium case, but large values for bottomonium (although orders 
of magnitude smaller than using ``Mass Input 1''). The situation improves if an unconfirmed state of 
bottomonium is included [namely, the $\eta_b(1S)$].
Let us note that, in all cases, using $V_{LGP+L}$ yields similar results
to the Cornell-potential case, with eigenstates slightly closer to the experimental values and with 
smaller errors.

The data obtained in the independent fits of charmonium and bottomonium spectra are then used to make a 
constrained fit, i.e.\ one with a common value for the string tension $\sigma$ of the two systems\footnote{We 
consider as input to this constrained fit just our second input method, i.e.\ column  ``Mass Input 2'' of Tables
\ref{tb:input_data_charm} and \ref{tb:input_data_bottom}, including the unconfirmed $\eta_b(1S)$ state.}.
Notice now that we will have three 
free parameters ($m_c$, $m_b$ and $\sigma$) and six states as inputs in the fit (the states 
1S, 2S and 1P of charmonium and bottomonium), resulting in three degrees of freedom. The results of this fit 
using the Cornell potential and $V_{LGP+L}$ are shown in Tables \ref{tb:results_simultaneous_fit_charm} 
and \ref{tb:results_simultaneous_fit_bottom}. 

\begin{figure*}
	\centering
\caption{Experimental mass spectrum for charmonium, together with the spin averages used as input in
our calculations. We also show our results in the $V_{LGP+L}$ and Cornell-potential cases. 
The fit shown is that of the constrained case and considering as input the states 1S, 2S and 1P of both systems.}
\label{fig:Mass_Spectrum_charm}
\includegraphics[width=\textwidth]{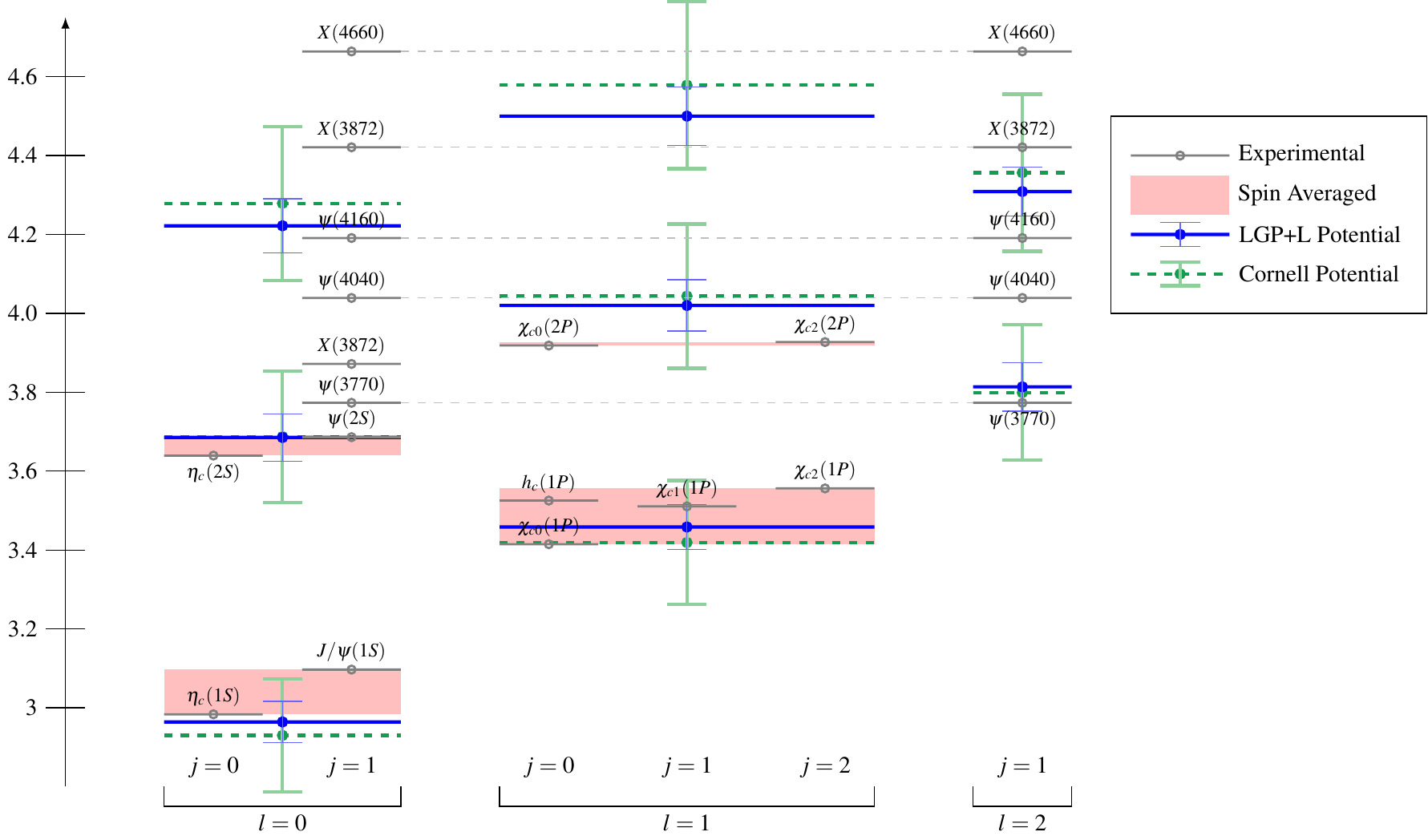}
\end{figure*}

\begin{figure*}
	\centering
\caption{Experimental mass spectrum for bottomonium, together with the spin averages used as input in
our calculations. We also show our results in the $V_{LGP+L}$ and Cornell-potential cases. 
The fit shown is that of the constrained case and considering as input the states 1S, 2S and 1P of both systems.}
\label{fig:Mass_Spectrum_bottom}%
\includegraphics[width=\textwidth]{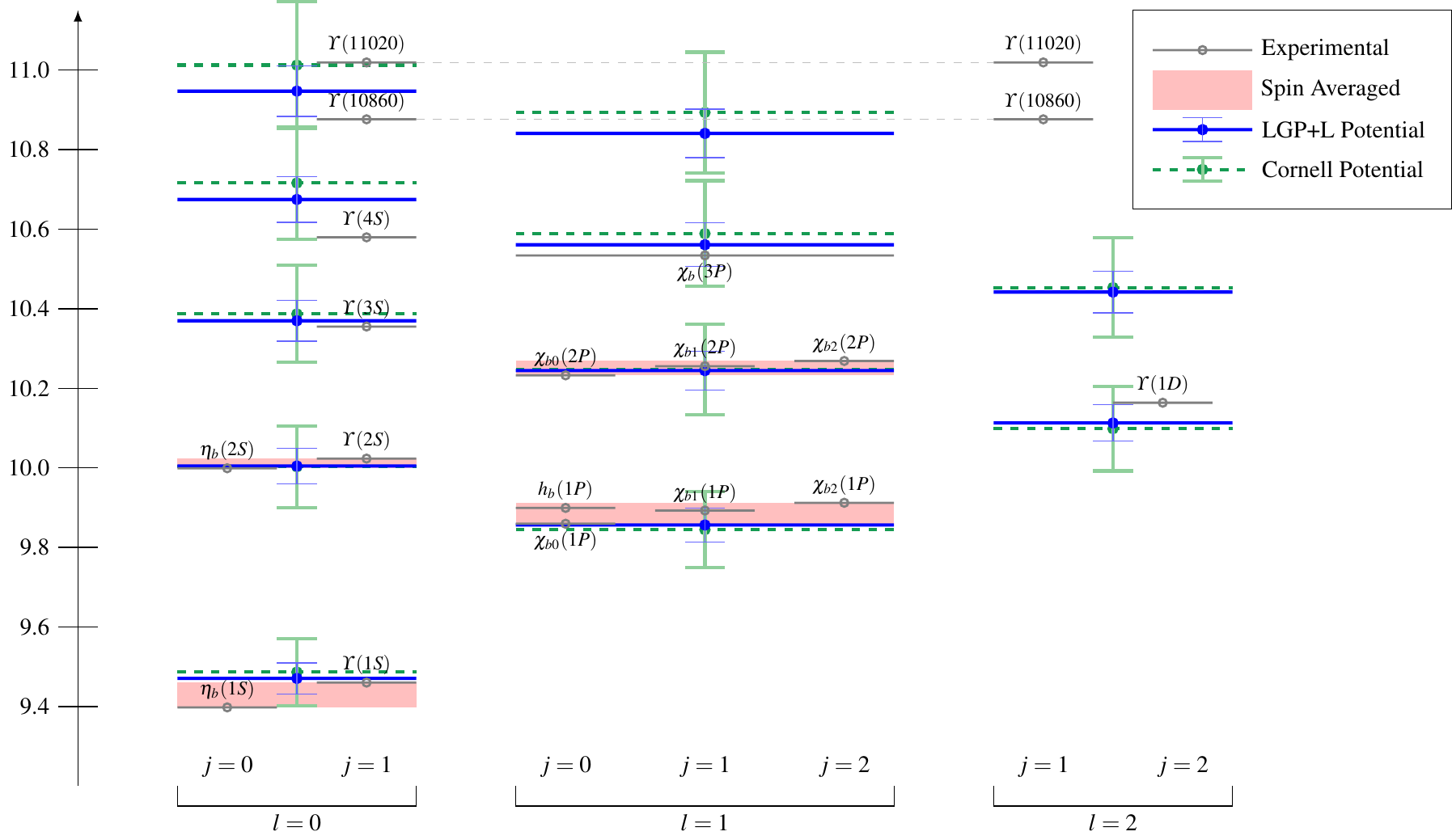}
\end{figure*}

In order to establish a confidence level for our parameters, we use the method described in detail in 
Ref.\ \cite{press2007numerical}, which consists in determining the region in parameter space for which 
$\chi^2$ increases by less than one unit with respect to its minimum value, for each of the parameters. 
The confidence level obtained for
each parameter is shown in Table \ref{tb:parameters_results_simultaneous_fit}. This also allows us to establish confidence 
levels for the eigenenergies.
In cases for which the obtained confidence level is asymmetric, we adopt the larger value as the error.
We then draw three random numbers, each one following a Gaussian distribution 
centered at the value of the optimal parameter found and with standard deviation equal to the symmetrized error. 
We use these numbers as parameters to compute the spectrum, repeating the process $N = 1000$ times.
The obtained values are presented in parentheses in Tables \ref{tb:results_simultaneous_fit_charm} and 
\ref{tb:results_simultaneous_fit_bottom}. 
The resulting potentials are shown for the charmonium case in Fig.\ \ref{fig:full_pot_comparison}.
A visual representation of the spectrum is provided in Figs.\ 
\ref{fig:Mass_Spectrum_charm} and \ref{fig:Mass_Spectrum_bottom}. We also compare these results with the ones 
in Ref.\ \cite{Chung:2008sm}, obtaining agreement with their eigenenergies to within $ \sim 10^{-3} $.

\begin{table}
\caption{Results for the charmonium eigenstates using $V_{LGP+L}$ and the Cornell potential 
in a constrained fit (see text).}
\centering
\label{tb:results_simultaneous_fit_charm}
\begin{tabular}{cS[table-format=2.5]S[table-format=2.5]S[table-format=2.5]S[table-format=2.5]}
\hline
\multicolumn{3}{c}{Charmonium Spectrum} \\ \hline
	\multicolumn{3}{c}{{$V_{LGP+L}$}} \\ \hline
{State}	& {Predicted mass}	& {Deviation from averaged}		\\
	& {(\si{\giga \electronvolt})}	& {spin state (\si{\giga \electronvolt})}	\\ \hline
{1S}	& 2.96 +- .05		& -0.10\\
{2S}	& 3.69 +- .06		& 0.01\\
{3S}	& 4.22 +- .07		& {-}\\
{1P}	& 3.46 +- .06		& -0.07\\
{2P}	& 4.02 +- .06		& 0.09\\
{3P}	& 4.50 +- .07		& {-}\\
{1D}	& 3.81 +- .06		& {-}\\
{2D}	& 4.31 +- .06		& {-}\\	\hline
\multicolumn{3}{c}{{Cornell Potential}} \\ \hline
{State}	& {Predicted mass}	& {Deviation from averaged}		\\
	& {(\si{\giga \electronvolt})}	& {spin state (\si{\giga \electronvolt})}	\\ \hline
{1S}	& 2.93 +- .14		& -0.14\\
{2S}	& 3.69 +- .17		& 0.01\\
{3S}	& 4.28 +- .20		& {-}\\
{1P}	& 3.42 +- .16		& -0.11	\\
{2P}	& 4.04 +- .18		& 0.12	\\
{3P}	& 4.58 +- .21		& {-}	\\
{1D}	& 3.80 +- .17		& {-}	\\
{2D}	& 4.36 +- .20		&	{-}\\	\hline

\end{tabular}
\end{table}

\begin{table}
\caption{Results for the bottomonium eigenstates using $V_{LGP+L}$ and the Cornell potential 
in a constrained fit (see text).}
\label{tb:results_simultaneous_fit_bottom}
\centering
\begin{tabular}{cS[table-format=2.5]S[table-format=2.5]S[table-format=2.5]S[table-format=2.5]}
\hline
\multicolumn{3}{c}{Bottomonium Spectrum} \\ \hline
\multicolumn{3}{c}{{$V_{LGP+L}$}}	\\ \hline
{State}	& {Predicted mass}	& {Deviation from averaged}		\\
	& {(\si{\giga \electronvolt})}	& {spin state (\si{\giga \electronvolt})}	\\ \hline
{1S}	& 9.47  +- .04		& 0.04 \\
{2S}	& 10.00 +- .04		& -0.01	\\
{3S}	& 10.37 +- .05		& 0.01\\
{4S}	& 10.67 +- .06		& 0.10\\
{5S}	& 10.95 +- .07		& {-}\\
{1P}	& 9.86	+- .04		& -0.03\\
{2P}	& 10.24 +- .05		& -0.01\\
{3P}	& 10.56 +- .05		& 0.03\\
{4P}	& 10.84 +- .06		& {-}\\
{1D}	& 10.11 +- .05		& -0.05\\
{2D}	& 10.44 +- .05		& {-}\\
{3D}	& 10.73 +- .06		& {-}\\ \hline	
\multicolumn{3}{c}{{Cornell Potential}} \\ \hline
{State}	& {Predicted mass}	& {Deviation from averaged}		\\
	& {(\si{\giga \electronvolt})}	& {spin state (\si{\giga \electronvolt})}	\\ \hline
{1S}	& 9.49 +- .08		& 0.06 \\
{2S}	& 10.00 +- .10		& -0.01\\
{3S}	& 10.39 +- .12		& 0.03\\
{4S}	& 10.72 +- .14		& 0.14\\
{5S}	& 11.01 +- .16		& {-}\\
{1P}	& 9.84  +- .10		& -0.04\\
{2P}	& 10.25 +- .11		& 0.00\\\
{3P}	& 10.59 +- .13		& 0.05\\
{4P}	& 10.89 +- .15		& {-}\\
{1D}	& 10.10 +- .11		& -0.06	\\
{2D}	& 10.45 +- .12		& {-}\\
{3D}	& 10.77 +- .14 		& {-}\\ \hline	
\end{tabular}
\end{table}

\begin{table}
\caption{Quark masses and string tension obtained from our fit. These parameters are used to 
obtain the spectrum in Tables \ref{tb:results_simultaneous_fit_charm} and \ref{tb:results_simultaneous_fit_bottom}.}
\label{tb:parameters_results_simultaneous_fit}
\centering
\resizebox{\columnwidth}{!}{%
\begin{tabular}{ccc} \hline
	$V_{LGP+L}$ & Cornell Potential & Quark Mass\\
	            &                   & in Ref.\ \cite{Agashe:2014kda} \\ \hline
	$m_c = \SI{1.16 +- 0.03}{\giga \electronvolt}$																& $m_c = \SI[parse-numbers = false]{1.11_{-0.02}^{+0.08}}{\giga \electronvolt}$ 		& $m_c = \SI{1.275 +- 0.025}{\giga \electronvolt}$\\[2mm]
	$m_b = \SI[parse-numbers = false]{4.61_{-0.01}^{+0.02}}{\giga \electronvolt}$	& $m_b = \SI[parse-numbers = false]{4.58_{-0.01}^{+0.04}}{\giga \electronvolt}$ 		& $m_b(\overline{MS}) = \SI{4.18 +- 0.03}{\giga \electronvolt}$\\[2mm]
	$\sigma = \SI{0.23 +- 0.01}{\giga \electronvolt}$															& $\sigma = \SI[parse-numbers = false]{0.26_{-0.03}^{+0.01}}{\giga \electronvolt}$	& $m_b(1S) = \SI{4.66 +- 0.03}{\giga \electronvolt}$\\[2mm]
	$\chi^2 = \num{6.20}$																													& $\chi^2 = \num{12.13}$																														&\\	\hline
\end{tabular}%
}
\end{table}

For both the charmonium and the bottomonium spectra, we obtain smaller errors and better agreement with
the spin-averaged experimental values in the $V_{LGP+L}$ case than in the Cornell-potential
one (see Tables \ref{tb:results_simultaneous_fit_charm} and \ref{tb:results_simultaneous_fit_bottom}).
This is especially true for the charmonium spectrum. In particular, the charmonium spectrum obtained
using the Cornell potential is not in agreement with experiment.
For bottomonium, instead, the comparison shows agreement.
This is not surprising, since bottomonium is a more closely nonrelativistic system.
Of course, it would be interesting to check if the inclusion of relativistic corrections
would allow a better description of the charmonium spectrum and higher accuracy in the bottomonium case. 
We point out that the errors in the $V_{LGP+L}$ case are of the same order of magnitude 
as in the spin-averaged experimental values for most states.

\begin{figure}
\centering
\caption{Comparison of the wave function $f(r)$ of the 1S state for bottomonium and charmonium using both 
potentials.}
 \label{fig:wave function}
	\includegraphics{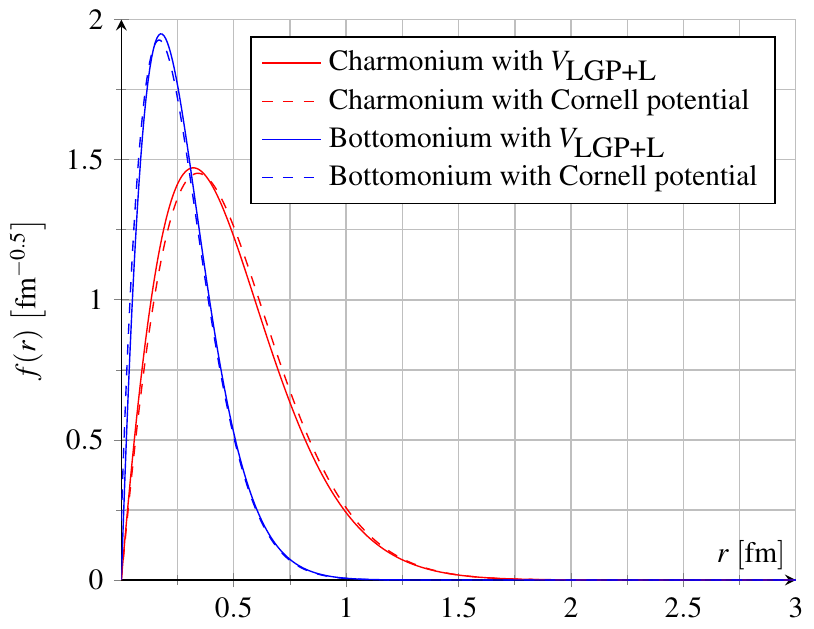}
\end{figure}

An advantage of our approach is that we have direct access to the radial wave function $f(r)$. 
We plot, as an example, the wave 
functions\footnote{The wave functions obtained using our code are not normalized. We interpolate the data 
and normalize $f(r)$ a posteriori.} for the 1S state for both potentials in the 
charmonium and bottomonium cases in Fig.\ \ref{fig:wave function}. 
We can see that the similarity between the two potentials (see Fig.\ \ref{fig:full_pot_comparison}) and the
obtained spectra is present for the wave functions as well. Also, note that the wave function is more 
extended for the charmonium states.

This direct access to the the wave function can be of interest in other applications, such as
effective field theories, for which one needs information on the typical distance between the 
quarks \cite{Brambilla:2014jmp}. We estimate this quantity by computing
\begin{equation}
d \;=\; 2 \langle r \rangle \;=\; 2 \int_0^\infty{r f(r)^2\, dr}\,. 
\label{eq:Typical_distance}
\end{equation}
We note that the factor $2$ in Eq.\ (\ref{eq:Typical_distance}) comes from the fact that $r$ corresponds to the 
distance between one of the quarks and the center-of-mass of the system. Some of these typical 
distances are presented in Tables \ref{tb:Typical_distances_charm} and \ref{tb:Typical_distances_bottom}.
\begin{table}
\caption{Typical interquark distances for charmonium. Errors are expected to be negligible.}
\label{tb:Typical_distances_charm}
\centering
\begin{tabular}{cS[table-format=1.2]S[table-format=1.2]}
\hline
\multicolumn{3}{c}{Charmonium} \\ \hline
{State}	& {$V_{LGP+L}$} 			& {Cornell Potential} \\ 
	& {distance (\si{\femto \metre})}	& {distance (\si{\femto \metre})} \\ \hline
{1S}	& 0.80					& 0.83\\
{2S}	& 1.59 					& 1.56\\
{3S}	& 2.20 					& 2.14\\
{1P}	& 1.29					& 1.28\\
{2P}	& 1.94					& 1.90\\
{3P}	& 2.50					& 2.43\\ \hline
\end{tabular}
\end{table}%
\begin{table}
\caption{Typical interquark distances for bottomonium. Errors are expected to be negligible.}
\label{tb:Typical_distances_bottom}
\centering
\begin{tabular}{cS[table-format=1.2]S[table-format=1.2]}
\hline
\multicolumn{3}{c}{Bottomonium} \\ \hline
{State}	& {$V_{LGP+L}$} 			& {Cornell Potential} \\ 
	& {distance (\si{\femto \metre})}	& {distance (\si{\femto \metre})} \\ \hline
{1S}	& 0.45					& 0.47\\
{2S}	& 0.94					& 0.94\\
{3S}	& 1.35					& 1.31\\
{1P}	& 0.76 					& 0.77\\
{2P}	& 1.18					& 1.16\\
{3P}	& 1.55 					& 1.49\\ \hline
\end{tabular}
\end{table}

Finally, we could also estimate decay widths, which are proportional to $|R(0)|^2$. Notice, however, that 
this calculation would require a more strict control of the numerical integration in the region near 
the origin, since the function $R(r) = f(r)/r$ typically shows a divergence for $r\to 0$.

\section{Conclusions}
\label{sec:Conclusions}

We briefly reviewed the potential-model approach for determining the spectrum of quarkonia and discussed
the simplest such approach, the Cornell potential. We then modified the procedure by replacing the free gluon 
propagator with one obtained using lattice simulations. The resulting $V_{LGP}$ potential is different from
the Coulomb-like potential, but is still non-confining. Inspection of Fig.\ \ref{fig:Comparison_Potentials}
shows that, up to the hadronic scale, the potential rises above zero, with 
a trend to rise further. This is no longer true for larger values of $r$, for which the potential
is damped.
In fact, in order to obtain a confining (linear) potential, the gluon propagator
should show a strong divergence, of $1/k^4$, in the infrared limit, as proven
in Ref.\ \cite{West:1982bt}.
Also, an oscillating behavior --- 
due to the complex poles of the lattice propagator \cite{Cucchieri:2011ig} --- is observed.

We therefore add a linear term to $V_{LGP}$, obtaining the 
$V_{LGP+L}$ potential in Eq.\ (\ref{eq:lgp+l_potential}).
We solve the associated Schr\"odinger equation numerically and compare our results with the spin-averaged 
spectrum. 
This is done for the Cornell potential and for $V_{LGP+L}$, both for charmonium and for bottomonium states.
The spectra computed using $V_{LGP+L}$ show a slight
improvement over the ones obtained using the Cornell potential, but no qualitative differences are observed. 
In particular, the resulting potentials are rather similar, as seen in Fig.\ \ref{fig:full_pot_comparison}.
A more accurate description of the spectra could 
be achieved by introducing relativistic corrections, as was done in Refs.\ \cite{Radford:2007vd,Olsson:1994cv}.

We were also able to obtain the wavefunction for all the states mentioned, which allows us to
estimate the interquark distance of the considered states. Let us note that
the wave functions are remarkably similar for the Cornell potential and $V_{LGP+L}$ 
(see Fig.\ \ref{fig:wave function}), even though the potentials are not identical
(see Fig.\ \ref{fig:full_pot_comparison}). This might suggest that the wave function is
somewhat insensitive to details of the potential.
In fact, a visual comparison between our wavefunctions and the one presented in 
\cite[Fig.\ 5]{Kawanai:2013aca} (corresponding to a different parametrization of the Cornell potential)
shows that they are also similar.

Let us mention that a study using a similar method was carried out in
Refs.\ \cite{Gonzalez:2011zc, Gonzalez:2012hx} to propose a potential 
for heavy-quarkonium states. In that case, the gluon propagator was taken from
a study of Schwinger-Dyson equations \cite{Aguilar:2008xm}.
This propagator is in qualitative agreement with the lattice results we
use. The main difference with respect to our study is that these authors
do not include the linear term in the potential, but consider an additive
contribution to the one-gluon-exchange potential, in such a way that the 
zero of the proposed potential coincides with the Cornell one.
This corresponds to fixing the self-energy of the static source \cite{Necco:2003jf},
which appears in higher-loop calculations. A constant term in the interquark
potential can also be interpreted as being related to the infrared divergence of the 
Fourier integral of a ``confining'' gluon propagator $1/k^4$ \cite{Lucha:1991vn}.
The spectrum obtained in \cite{Gonzalez:2012hx} is in general agreement with the expected values.
As discussed here, this procedure is not able 
to generate a linearly rising potential, 
associated with confinement in the static case (see also \cite{Vento:2012wp}). 
In our study, we have
ensured that the expression in Eq.\ (\ref{eq:lattice_gluon_propagator}) approaches 
$1/k^2$ in the ultraviolet limit by rescaling the corresponding term in the 
potential by a suitable constant.

We note again that our aim was to gain a qualitative understanding of the interplay 
between perturbative and nonperturbative features of the interquark potential. As
verified in our study, even though the full nonperturbative gluon propagator was
used, the potential is non-confining, i.e.\ confinement is washed away by the use of 
the (tree-level) perturbative approximation for the interaction. Nevertheless, the 
resulting potential (with the addition of a linear term) provides a slightly better
description of the spectra, with the same number of fit parameters as the Cornell potential.

\vskip 3mm
The authors thank B. Blossier and F.\ Navarra for useful comments. W.S.\ thanks the Brazilian funding agencies
FAPESP and CNPq for financial support. A.C.\ and T.M.\ thank CNPq for partial support.

\bibliographystyle{spphys}
\bibliography{biblio2}
\end{document}